# Quantum Physics, Algorithmic Information Theory and the Riemann's Hypothesis

R. V. Ramos

rubens.viana@pq.cnpq.br

*Lab. of Quantum Information Technology, Department of Teleinformatic Engineering – Federal University of Ceara - DETI/UFC, C.P. 6007 – Campus do Pici - 60455-970 Fortaleza-Ce, Brazil.*

In the present work the Riemann's hypothesis (RH) is discussed from four different perspectives. In the first case, coherent states and the Stenger's approximation to Riemann-zeta function are used to show that RH avoids an indeterminacy of the type 0/0 in the inner product of two coherent states. In the second case, the Hilber-Pólya conjecture with a quantum circuit is considered. In the third case, randomness, entanglement and the Möbius' function are used to discuss the RH. At last, in the fourth case, the RH is discussed by inverting the first derivative of the Chebyshev function. The results obtained reinforce the belief that the RH is true.

Keywords: Riemann-Zeta function, coherent states, randomness, Möbius function, Chebyshev function.

## 1. Introduction

The Riemann hypothesis (RH) is an exciting problem that has challenged many scientists. Basically, the Riemann's hypothesis states that all the non-trivial zeros of the zeta function $\zeta(z)$, a function of a complex variable $z$ that analytically continues the sum of the infinite series $\Sigma_n(1/n^z)$, are of the form $z = \frac{1}{2} \pm it$ (the trivial zeros are the negative even integers). In particular, it has established a bridge between physics and number theory, and now several physical systems related to the Riemann-zeta function are known [1-6]. In general, scientists look for a physical reason that forbids the existence of zeros that do not lie on the critical line, $\text{Re}(z) = \frac{1}{2}$.

In this work, the RH is discussed from four different perspectives. Firstly, using coherent states and the Stenger's approximation to Riemann-zeta function [7], it is shown that the real part of the nontrivial zeros of the Riemann-zeta function should be ½ in order to avoid an indeterminacy of the type 0/0 in the inner product of two coherent states. In order to show this, a quantum operator based on the Stenger's function is introduced. Secondly, the Hilber-Pólya conjecture is considered by constructing a unitary matrix (a quantum circuit) related to any finite amount of zeros on the critical line. Then, a quantum state based on that unitary matrix is introduced. It is shown that the non-trivial zeros out of the critical line, if they exist, have to obey a condition, otherwise it would imply in an inappropriate behavior of the proposed quantum state. In the third perspective, the randomness of a bit string obtained from a series of integer number is linked to the entanglement of a quantum state: The larger the randomness the larger is the entanglement. Then it is shown that using the Möbius function a maximally entangled state is obtained implying that the binary string obtained by the Möbius function is incompressible what, according to the algorithmic information theory, implies in an equal probability for 1's and 0's in that binary sequence. At last, the first derivative of the Chebyshev function is considered. Basically, the first derivative of the Chebyshev function is zero everywhere except for spikes at the values of the argument that are prime powers, therefore, a binary sequence can be constructed using it. It is argued that it is possible to make the inverse path. Having the binary sequence with 1's only at the prime powers positions, the zeros can be (numerically) obtained by an optimization procedure.

This work is outlined as follows: In Section 2, the RH is discussed by using coherent states and the Stenger's approximation to Riemann-zeta function. In Section 3, the Hilbert-Pólya conjecture is considered using a quantum circuit. In Section 4, randomness, entanglement and the Möbius' function are used to discuss the RH. In Section 5, the RH is discussed by inverting the first derivative of the Chebyshev function. At last, the conclusions are drawn in Section 6.

## 2. Coherent states, the Stenger function and the zeros of the Riemann-zeta function

Initially, consider the following function [7]

$$F(z) \equiv \left(1 - \frac{2}{2^z}\Gamma(z)\right)\zeta(z) = \int_0^\infty \frac{t^{z-1}}{e^t + 1} dt. \tag{1}$$

The function $F(z)$ has the same zeros as the Riemann-zeta function $\zeta(z)$ in the critical strip $d = \{z \in \mathbb{C} : 0 < \mathbb{R}(z) < 1\}$. Furthermore, as shown in [7], the function $F_h(x+iy)$, here named as Stenger function, given by

$$F_h(x+iy) = \begin{cases} \sum_{k=-\infty}^{\infty} \frac{e^{kh(x+iy)}}{\exp(e^{kh}) + 1} & \text{if } |y| < \pi/h \\ 0 & \text{if } |y| > \pi/h \end{cases} \tag{2}$$

converges to $F(z)$ at all points of the critical strip $d$.

In order to use the coherent states, initially eq. (2) is rewritten by changing the complex variable $z = x + iy$ by the annihilation operator $a$, hence, a new operator $\hat{F}_h$ is defined as

$$\hat{F}_h = \sum_{k=-\infty}^{\infty} \frac{e^{kha}}{\exp(e^{kh})+1} \tag{3}$$

Since $a|\alpha\rangle = \alpha|\alpha\rangle$, where $|\alpha\rangle$ is a coherent state and $\alpha$ is a complex number, one has that

$$\hat{F}_h|\alpha\rangle = \sum_{k=-\infty}^{\infty} \frac{e^{kha}}{\exp(e^{kh})+1}|\alpha\rangle = \sum_{k=-\infty}^{\infty} \frac{e^{kh\alpha}}{\exp(e^{kh})+1}|\alpha\rangle = F_h(\alpha)|\alpha\rangle. \tag{4}$$

Now, let us remind three important properties of coherent states:

1) $D(\alpha)|\beta\rangle = \exp(i\mathrm{Im}(\alpha\beta^*))|\alpha+\beta\rangle$, where $D(.)$ is the Glauber displacement operator.
2) $|\langle\beta|\alpha\rangle|^2 = \exp(-|\alpha-\beta|^2)$, hence, $|\langle\beta|\alpha\rangle|^2 > 0$ if $\alpha$ and $\beta$ have finite values.
3) Both operators $\hat{F}_h(.)$ and $D(.)$ map coherent states to coherent states.

Calculating the inner product of the normalized versions of the two coherent states $\hat{F}_h D(\alpha_j)|\alpha_j^*\rangle$ and $\hat{F}_h D(\alpha_j)|0\rangle$, where $\alpha_j$ and $\alpha_j^*$ are conjugated zeros of the Riemann-zeta function, $\zeta(\alpha_j) = \zeta(\alpha_j^*) = 0$, one gets

$$\delta_h = \frac{\left|\langle 0|D^\dagger(\alpha_j)\hat{F}_h^\dagger \hat{F}_h D(\alpha_j)|\alpha_j^*\rangle\right|}{\sqrt{\langle\alpha_j^*|D^\dagger(\alpha_j)\hat{F}_h^\dagger \hat{F}_h D(\alpha_j)|\alpha_j^*\rangle}\sqrt{\langle 0|D^\dagger(\alpha_j)\hat{F}_h^\dagger \hat{F}_h D(\alpha_j)|0\rangle}} =$$
$$\left|\frac{F_h^*(\alpha_j)F_h(2\mathrm{Re}(\alpha_j^*))}{F_h^*(\alpha_j)F_h(2\mathrm{Re}(\alpha_j^*))}\right|\left|\langle\alpha_j|2\mathrm{Re}(\alpha_j^*)\rangle\right| \tag{5}$$

Equation (5) is a trivial result when $\alpha_j$ is not a zero of $F_h$. However, since $F_h$ in (2) converges to $F$ in (1) when $h$ tends to zero and $F$ and the Riemann-zeta function have the same zeros on the critical strip, one has that $F_h(\alpha_j)$ goes to zero when $h$ tends to zero. Since $|\langle\alpha_j|2\mathrm{Re}(\alpha_j^*)\rangle|$ in (5) is larger than zero and finite, the condition required for (5) do not have an indeterminacy of the type 0/0 when $h$ goes to zero is

$$\lim_{h\to 0} F_h(2\mathrm{Re}(\alpha_j)) = \infty. \tag{6}$$

This happens only for $2\mathrm{Re}(\alpha_j) = 1$ (for the analytical continuation of the Riemann-zeta function only $\zeta(1)$ is not defined), hence $\mathrm{Re}(\alpha_j) = 1/2$. Hence, it was shown that in order to avoid an indeterminacy of the type 0/0 in the inner product of two coherent states, the real part of the non-trivial zeros should be equal to ½.

## 3. Riemannian quantum circuit

According to Hilbert and Pólya discussion, the zeros of the Riemann-zeta function could be the spectrum of an operator $R = I/2 + iH$, where $H$ is self-ajoint and interpreted as a Hamiltonian. There are several physical systems related to the zeros of the Riemann-zeta function [1]. It is interesting to note that, using $R$, it is always possible to construct a physical system, in an arbitrary finite dimensional space, related to the zeros [5]. Firstly, one constructs the following unitary matrix

$$U_R = R^\dagger/R = [0.5I - iH]/[0.5I + iH]. \qquad (7)$$

Since $H$ is Hermitean, one has the decomposition $H = UDU^\dagger$ where $D$ is a diagonal matrix whose entries are the eigenvalues of $H$ while $U$ is a unitary matrix whose columns are the eigenvalues of $H$. Hence,

$$U_R = R^\dagger/R = U\frac{[I/2 - iD]}{[I/2 + iD]}U^\dagger. \qquad (8)$$

Since the unitary operation does not change the eigenvalues, the following unitary matrix has also the eigenvalues linked to the zeros of the Riemann-zeta function

$$U'_R = \frac{[I/2 - iD]}{[I/2 + iD]} = \left[1/2\begin{bmatrix}1 & 0 & 0\\ 0 & \ddots & 0\\ 0 & 0 & 1\end{bmatrix} - i\begin{bmatrix}s_1 & 0 & 0\\ 0 & \ddots & 0\\ 0 & 0 & s_n\end{bmatrix}\right] \Big/ \left[1/2\begin{bmatrix}1 & 0 & 0\\ 0 & \ddots & 0\\ 0 & 0 & 1\end{bmatrix} + i\begin{bmatrix}s_1 & 0 & 0\\ 0 & \ddots & 0\\ 0 & 0 & s_n\end{bmatrix}\right] \Rightarrow$$

$$U'_R = \begin{bmatrix}\frac{1/2 - is_1}{1/2 + is_1} & 0 & 0\\ 0 & \ddots & 0\\ 0 & 0 & \frac{1/2 - is_n}{1/2 + is_n}\end{bmatrix}. \qquad (9)$$

Taking $z_j = 1/2 + is_j$ as the $j$-th zero on the critical line (with imaginary part positive), the following unitary matrix represents a physical system that can be implemented as a quantum circuit [8], related to $n$ zeros of the Riemann-zeta function:

$$U_{Riemann} = U \begin{bmatrix} \dfrac{1/2 - is_1}{1/2 + is_1} & 0 & 0 \\ 0 & \ddots & 0 \\ 0 & 0 & \dfrac{1/2 - is_n}{1/2 + is_n} \end{bmatrix} U^\dagger = U \begin{bmatrix} \dfrac{z_1^*}{z_1} & 0 & 0 \\ 0 & \ddots & 0 \\ 0 & 0 & \dfrac{z_n^*}{z_n} \end{bmatrix} U^\dagger. \qquad (10)$$

Since $U$ in (10) is arbitrary, there are an infinite number of physical systems related to the zeros. Furthermore, the zeros used in (10) do not need to be consecutive. Thus, one can take any finite number of zeros, anywhere in the critical line, and build a quantum circuit whose eigenvalues are related to them in a very clear way: each eigenvalue depends on only one zero. In other words, all zeros of the Riemann-zeta function that lie on the critical line are related to a physical system that can be (at least in principle) constructed with any technology useful for construction of quantum computers.

Now, consider the following quantum state belonging to a Hilbert space with dimension $K$:

$$|\psi_n\rangle = N_n \left[ |x_0\rangle + |x_1\rangle + \ldots + |x_{n-1}\rangle \right]. \qquad (11)$$

In (11) $N_n$ is the normalization constant and in general $|x_i\rangle$ and $|x_j\rangle$ are not orthogonal. The states in the superposition are related by $|x_{i+1}\rangle = U|x_i\rangle$, where $U$ is a unitary operation. Therefore, eq. (11) can be rewritten as

$$|\psi_n\rangle = N_n \left[ U^0 |x_0\rangle + U^1 |x_0\rangle + \ldots + U^{n-1} |x_0\rangle \right] = N_n \left[ U^0 + U^1 + \ldots + U^{n-1} \right] |x_0\rangle. \qquad (12)$$

Now, using the spectral decomposition of $U$

$$U = \sum_{k=1}^{K} \lambda_k |\lambda_k\rangle \langle \lambda_k| \qquad (13)$$

in (12), one gets

$$|\psi_n\rangle = N_n \left[\sum_{l=0}^{n-1}\sum_{k=1}^{K}\lambda_k^l |\lambda_k\rangle\langle\lambda_k|\right]|x_0\rangle = N_n \sum_{k=1}^{K}\left[\sum_{l=0}^{n-1}\lambda_k^l\right]\langle\lambda_k|x_0\rangle|\lambda_k\rangle =$$

$$N_n \sum_{k=1}^{K}\left(\frac{1-\lambda_k^n}{1-\lambda_k}\right)\langle\lambda_k|x_0\rangle|\lambda_k\rangle. \qquad (14)$$

Substituting the eigenvalues $\lambda_k = \exp(i\phi_k)$ in (14), one obtains

$$|\psi_n\rangle = N_n \sum_{k=1}^{K}\langle\lambda_k|x_0\rangle\left(\frac{1-e^{in\phi_k}}{1-e^{i\phi_k}}\right)|\lambda_k\rangle = \sum_{k=1}^{K} c_k^n |\lambda_k\rangle. \qquad (15)$$

The probabilities $|c_k^n|^2$ are given by

$$|c_k^n|^2 = |N_n|^2 |\langle\lambda_k|x_0\rangle|^2 \left(\frac{1-\cos(n\phi_k)}{1-\cos(\phi_k)}\right) = |N_n|^2 |\langle\lambda_k|x_0\rangle|^2 \frac{\sin^2(n\phi_k/2)}{\sin^2(\phi_k/2)}. \qquad (16)$$

Now, let us consider the measurement represented by the Hermitean operator

$$H = \sum_{k=1}^{K}\sin^2\left(\frac{\phi_k}{2}\right)|\lambda_k\rangle\langle\lambda_k|. \qquad (17)$$

where $\phi_k \in [0, \pi/2]$. Thus, the average value of the outcome of the measurement of state (15) using the operator (17) is given by

$$\langle\psi_n|H|\psi_n\rangle = \sum_{k=1}^{K}\sin^2\left(\frac{\phi_k}{2}\right)|c_k^n|^2 = |N_n|^2 \sum_{k=1}^{K}|\langle\lambda_k|x_0\rangle|^2 \sin^2(n\phi_k/2). \qquad (18)$$

Choosing $|x_0\rangle = (1/K^{1/2})[|\lambda_1\rangle+|\lambda_2\rangle+\ldots+|\lambda_K\rangle]$, one has $|\langle\lambda_k|x_0\rangle|^2 = 1/K$ for all $k$ values and, hence, (18) can finally be written as

$$\langle\psi_n|H|\psi_n\rangle = \frac{|N_n|^2}{K}\sum_{k=1}^{K}\sin^2(n\phi_k/2) = \frac{|N_n|^2}{2K}\sum_{k=1}^{K}[1-\cos(n\phi_k)]. \qquad (19)$$

The behavior of $\langle\psi_n|H|\psi_n\rangle$ depends on the values of $\phi_k$'s and $n$. One may also note that

$$|\psi_{n-1}\rangle = N_{n-1}\left[|x_0\rangle + |x_1\rangle + \ldots + |x_{n-2}\rangle\right] \Rightarrow U|\psi_{n-1}\rangle = \frac{N_{n-1}}{N_n}\left(|\psi_n\rangle - N_n|x_0\rangle\right) \quad (20)$$

$$[U,H] = 0 \Rightarrow \langle\psi_n|U^\dagger H U|\psi_n\rangle = \langle\psi_n|H|\psi_n\rangle \quad (21)$$

hence,

$$\langle\psi_{n-1}|H|\psi_{n-1}\rangle = \langle\psi_{n-1}|U^\dagger H U|\psi_{n-1}\rangle =$$
$$|N_{n-1}|^2\left(\frac{\langle\psi_n|H|\psi_n\rangle}{|N_n|^2} + \langle x_0|H|x_0\rangle - \frac{\langle\psi_n|H|x_0\rangle}{N_n^*} - \frac{\langle x_0|H|\psi_n\rangle}{N_n}\right). \quad (22)$$

From (22) one can see that, in general,

$$\frac{\langle\psi_n|H|\psi_n\rangle}{|N_n|^2} \neq \frac{\langle\psi_{n-1}|H|\psi_{n-1}\rangle}{|N_{n-1}|^2}. \quad (23)$$

Now, coinsidering $U = U_{\text{Riemann}}$ ($K$ x $K$) in (13) one has $\phi_k = arg(z_k^*/z_k) = -2tan^{-1}(2s_k)$. Using the approximation $\phi_k \approx \pi - 1/s_k$ in (19), one gets

$$\frac{\langle\psi_n|H|\psi_n\rangle}{|N_n|^2} = \frac{1}{2K}\sum_{k=1}^{K}[1-\cos(n\phi_k)] = \frac{1}{2K}\sum_{k=1}^{K}\left[1-(-1)^n\cos\left(n\frac{1}{s_k}\right)\right]. \quad (24)$$

From (24) one can see that, if $\cos(n/s_k) = 0$ for all $n$ values then $\langle\psi_n|H|\psi_n\rangle/|N_n|^2 = 1/2$. In this case one would have $\langle\psi_n|H|\psi_n\rangle/|N_n|^2 = \langle\psi_{n-1}|H|\psi_{n-1}\rangle/|N_{n-1}|^2$, that is in disagreement with (23). In order to avoid that, the following condition should be satisfied

$$\frac{n}{s_k} \neq (2m+1)\frac{\pi}{2} \quad \forall\, k. \tag{25}$$

Thus, the values $s_k = [2n/(2m+1)]/\pi$ are forbidden. In fact, one can see in [9] that the imaginary part of the zeros on the critical line does not have such appearance.

On the other hand, if there are zeros out of the critical line and such zeros are used to construct $U$ (if possible), then (24) would be rewritten as

$$\frac{\langle \psi_n | H | \psi_n \rangle}{|N_n|^2} = \frac{1}{2K}\sum_{k=1}^{K}\left[1-(-1)^n \cos\left(n\frac{1+2\varepsilon_k}{s_k}\right)\right]. \tag{26}$$

In (26) $-\tfrac{1}{2} < \varepsilon_k < \tfrac{1}{2}$. Now, in order to avoid a monotonous behavior of (26), the condition to be satisfied is

$$\frac{n(1+2\varepsilon_k)}{s_k} \neq (2m+1)\frac{\pi}{2} \quad \forall\, k. \tag{27}$$

Thus, the values $\varepsilon_k = ((2m+1)/n)(\pi/4)s_k - 1/2$ are forbidden.

## 4. Entanglement and the randomness of bit strings originated from sequences of integer numbers

Firstly, consider that $S$ is a finite sequence of integer numbers: $S = \{s_1, s_2, \ldots, s_n\}$. Now, let us assume there is a question $Q$ about each element of $S$ whose answer is true or false. Thus, one can build the binary string $B = Q(s_1)Q(s_2)\ldots Q(s_n)$, where $Q(s_i) =$ '1' if $s_i$ satisfies the question $Q$, otherwise $Q(s_i) =$ '0'. The subset of $S$ formed by the elements of $S$ that satisfy $Q$ is the set $S_Q$. The randomness of the binary string $B$ depends on the question $Q$. Two examples of questions are: 1) which elements of $S$ are even? 2) Which elements of $S$ are primes? Chaitin linked the randomness of a bit string to the length of the shortest computer program able to generate it [10,11]. A given bit string is said to have high randomness if it has (roughly) the same length of the shortest computer program able to generate it. In this case, such bit string is said to be incompressible. This idea is here explored by using a simple quantum computer model.

Given the (not normalized) quantum state $|S\rangle = \sum_{i=1}^{n}|s_i\rangle$ (related to the set of integer numbers $S = \{s_1, s_2, \ldots, s_n\}$) and the question $Q$, the quantum computer composed by a unitary operation $U_Q$ and the quantum state $|W\rangle = \sum_{i=1}^{m}|w_i\rangle$ (related to the set of integer numbers $W = \{w_1, w_2, \ldots, w_m\}$), is used to mark the elements of $|S\rangle$ that satisfy $Q$: $U_Q|s_i\rangle|w_j\rangle|0\rangle = |s_i\rangle|w_j\rangle|Q(s_i, w_j)\rangle$. Thus, $W$ is the set of integer numbers, with the minimum cardinality, required to answer the question $Q$ for all elements of $S$. If $U_Q|s_i\rangle|w_j\rangle|0\rangle = |s_i\rangle|w_j\rangle|1\rangle$ then $w_j$ is named a witness of $s_i$. The quantum operation that marks all the elements of $S$ that satisfy $Q$ is

$$|B\rangle = U_Q \frac{1}{\sqrt{2^{n+m}}} \left[|s_1\rangle + |s_2\rangle + \ldots + |s_n\rangle\right]\left[|w_1\rangle + |w_2\rangle + \ldots + |w_m\rangle\right]|0\rangle$$
$$= \frac{1}{\sqrt{2^{n+m}}} \sum_{i=1}^{n}\sum_{j=1}^{m}|s_i\rangle|w_j\rangle|Q(s_i, w_j)\rangle. \quad (28)$$

One may note that having the quantum state (28), quantum search [12] can be used to produce a quantum state whose elements are only the elements of $S_Q$, as well a quantum counting algorithm [12] can be used to determine the cardinality of $S_Q$.

Regarding the randomness defined by Chaitin, the quantum state $|W\rangle$ plays the role of the classical computer program. The quantum state $|W\rangle$ has the required information to generate $|B\rangle$ using $U_Q$ (the quantum state $|S\rangle$ is the input of the quantum computer in the same way that $S$ is the input of the algorithm $C$ in a classical computer). If $|W| = m < q = |S_Q|$, where $|x|$ means the cardinality of the set $x$, then the binary sequence $B$ obtained from $S$ and $Q$ is compressible. The larger the difference $q-m$ the larger is the compression. On the other hand, if $q = m$ (both sets have the same cardinality) then the binary sequence $B$ is incompressible and it has maximum randomness.

The bit sequence $B$ can be compressed if there is at least one element of $W$ that is a witness of at least two elements of $S_Q$. For example, $U_Q|s_i\rangle|w_j\rangle|0\rangle = |s_i\rangle|w_j\rangle|1\rangle$ and $U_Q|s_k\rangle|w_j\rangle|0\rangle = |s_k\rangle|w_j\rangle|1\rangle$. Hence, $w_j$ is witness of $s_i$ and $s_k$. On the other hand, if each element of $S_Q$ has a unique witness, then the number of witnesses is equal to the cardinality of $S_Q$ and the bit sequence $B$ is incompressible. At last, in order to have the set $W$ with the smallest cardinality, an element of $S$ should not have two (or more) different witnesses. For example, if $U_Q|s_i\rangle|w_j\rangle|0\rangle = |s_i\rangle|w_j\rangle|1\rangle$ and $U_Q|s_i\rangle|w_k\rangle|0\rangle = |s_i\rangle|w_k\rangle|1\rangle$ then $w_j$ and $w_k$ are both witnesses of $s_j$. In this case, one has to check if of them can be discarded and the cardinality of $W$ decreased by one unity.

Thus, the present work tries to answer the following question: given the (not normalized) states $|S\rangle = \sum_{i=1}^{n}|s_i\rangle$ and $|S_Q\rangle = \sum_{i=1}^{l}|s_{Qi}\rangle$, where each $s_{Qi}$ is an element of $S$ that satisfies a given question $Q$, how much information is required by the quantum circuit that realizes the operation $U|S\rangle = |S_Q\rangle$? In order to get this answer, we divided $U$ in

two parts: a unitary operation $U_Q$ and a (not normalized) quantum state $|W\rangle = \sum_{i=1}^{m}|w_i\rangle$. The last carries the amount of information required by $U$. Since $|W\rangle$ is linked to the question $Q$ and, hence, to $|S_Q\rangle$, the randomness, that depends on $Q$, depends also on the relation between the cardinalities of $W$ and $S_Q$ or, equivalently, depends on the relation between the number of terms of $|W\rangle$ and $|S_Q\rangle$.

As written before, using quantum search to select only the marked states ($Q(s_i,w_j) = 1$) in (28) one can get the quantum state whose elements are only the elements of $S_Q$, $|S_Q\rangle$ and their witnesses. Now, let us assume that $|S_Q| = l$. Considering the randomness one has three possibilities for the state $|S_Q\rangle$:

$$I)\,|S_Q\rangle = \frac{1}{\sqrt{2^l}}\left[|s_{Q1}\rangle+|s_{Q2}\rangle+...+|s_{Ql}\rangle\right]|w\rangle \qquad (29)$$

$$II)\,|S_Q\rangle = \frac{\left[|s_{Q1}\rangle+...+|s_{Qk}\rangle\right]}{\sqrt{2^k}}|w_1\rangle + \frac{\left[|s_{Qk+1}\rangle+...+|s_{Ql}\rangle\right]}{\sqrt{2^{l-k}}}|w_2\rangle \qquad (30)$$

$$III)\,|S_Q\rangle = \frac{1}{\sqrt{2^l}}\left[|s_{Q1}\rangle|w_1\rangle+|s_{Q2}\rangle|w_2\rangle+...+|s_{Ql}\rangle|w_l\rangle\right]. \qquad (31)$$

Equation (29) represents the situation in which there is no randomness, the compression is maximal. Equation (30) (that is a particular situation used to make easier the understanding), by its turn, represents a situation where there is some randomness and, hence, the compression is not maximal. In this example there are only two witnesses, thus the set $S_Q$ is partitioned in two parts. The extension for larger partitions is straightforward. At last, equation (31) represents the situation where there is maximal randomness and none compression. Observing (29)-(31) one can note a relation between the randomness and the entanglement between the quantum states of $|S_Q\rangle$ and $|W\rangle$. In equation (29) the knowledge of the value of $|w\rangle$ gives none information about the value of $|S_Q\rangle$ (any $S_{Qi}$ value is equally probable for $i = 1,…,l$). In equation (30), the knowledge of $|w\rangle$ gives some information about the value of $|S_Q\rangle$ (for example, if $w = w_1$ then $S_Q$ belongs to the set $[S_{Q1},…,S_{Qk}]$). Finally, in equation (31), the knowledge of the value of $|w\rangle$ gives complete information about the value of $|S_Q\rangle$ (for example, if $w = w_i$ then $S_Q = S_{Qi}$).

As examples, let us start by considering $S$ the integer numbers from 1 up to $n$. The question $Q$ is: which elements of $S$ belong to the sequence $x_{k+1}= px_k+q$, with $x_0=1$? In this case $U_Q$ implements the operation $U_Q|s\rangle|w\rangle|0\rangle = |s\rangle|w\rangle|((s \mod p)\oplus w)\oplus 1\rangle$ (for simplification it is understood that $(s \mod p)\oplus w = 0$ if $(s \mod p) = w$) and the set $W$ is only $W = \{q\}$, what means there is no randomness and the compression is maximal.

Now, we consider the randomness in the prime distribution. The set $S$ is the set of integer numbers from 2 to $n$ (where $n$ is an arbitrarily large number), and the question is:

Which elements of $S$ are not prime numbers? The elements of $W$ will be witnesses of composite numbers. Thus, the elements of the set $W$ are the prime numbers from 2 up to $n^{1/2}$. The quantum operation is $U_Q|s\rangle|w\rangle|0\rangle = |s\rangle|w\rangle|(\text{isinteger}(s/w))\rangle$, where the function isinteger($x$) returns '1' ('0') if $x$ is (not) an integer number. Since the number of primes between 2 and square root of $n$ is smaller than the number of composite numbers between 1 and $n$, there is some compression. However, the minimal length of $W$ is $\pi(n^{1/2})$ where $\pi$ is the prime count function, thus, as expected, there is some randomness in the prime distribution.

A more complicate situation occurs when one considers the Möbius function: $\mu(k) = 0$ if $k$ has at least one repeated prime factor, $\mu(1) = 1$ and $\mu(k) = (-1)^l$ when $k$ is a product of $l$ distinct primes. The Möbius function is very important in number theory, for example, it is related to the Riemann hypothesis: if the probability of $\mu(k) = +1$ (or -1) is 50% then the Riemann hypothesis is true. Let $S = \{s_1,..,s_n\}$ be the set composed by the first $n$ (again we consider $n$ an arbitrary large number) integer numbers with Möbius function different of zero, $\mu(s_i) = \pm 1$ for $i \in [1,n]$. The question $Q$ is: Which elements of $S$ have Möbius function equal to +1? The bit sequence originate from $S$ and $Q$ is $B = \{0.5[1+\mu(s_1)], 0.5[1+\mu(s_2)], …, 0.5[1+\mu(s_n)]\}$. If this sequence is incompressible (or maximally random, what happens when prob($\mu(k) = 1$) = prob($\mu(k) = -1$) =50%) then the Riemann hypothesis is true.

There are two possibilities for the witnesses: 1) The set of numbers having $\mu = +1$. In this case the set of witnesses is the same set of numbers to be tested and the $U_Q$ operation performs like

$$U_Q|s_{Qi}\rangle|w_i\rangle|0\rangle = |s_{Qi}\rangle|w_i\rangle|\overline{(s_{Qi} \oplus w_i)}\rangle. \tag{32}$$

Hence, for each $S_{Qi}$ there is a $w_i = S_{Qi}$ and there is no compression. 2) The set of numbers having $\mu = -1$. In this case, the operation $U_Q$ would be

$$U_Q|s_{Qi}\rangle|w_i\rangle|0\rangle = |s_{Qi}\rangle|w_i\rangle|\text{isprime}(s_{Qi}/w_i)\rangle. \tag{33}$$

In (33) the function isprime($k$) returns the value 1 (0) if $k$ is (not) a prime number. In fact, for any integer $k$ with $\mu(k) = 1$ there exist an integer $t$ with $\mu(t) = -1$ such that $k/t$ is prime. For example: let $p_1,…,p_{l-1},p_l$ be $l$ distinct prime numbers, where $l$ is an even number. Then $t = p_1 p_2 … p_{l-1}$ ($\mu(t) = -1$) is a witness that $\mu(k) = 1$ for $k = p_1 p_2 … p_l$, since $k/t = p_l$. Now, let us assume the following situation

$$U_Q |s_{Qi}\rangle |w_i\rangle |0\rangle = |s_{Qi}\rangle |w_i\rangle |1\rangle \quad (34)$$

$$U_Q |s_{Qj}\rangle |w_i\rangle |0\rangle = |s_{Qj}\rangle |w_i\rangle |1\rangle \quad (35)$$

$$U_Q |s_{Qk}\rangle |w_k\rangle |0\rangle = |s_{Qk}\rangle |w_k\rangle |1\rangle \quad (36)$$

$$U_Q |s_{Ql}\rangle |w_k\rangle |0\rangle = |s_{Ql}\rangle |w_k\rangle |1\rangle \quad (37)$$

Hence, $w_i$ is witness of $S_{Qi}$ and $S_{Qj}$, while $w_k$ is witness of $S_{Qk}$ and $S_{Ql}$. In this case $w_j$ and $w_l$ could in principle be discarded, indicating a compression. However, if they are discarded the element of $S_Q$ equal to $w_j.w_l$ would stay without witnesses, what is not acceptable. Therefore, $w_j$. and $w_l$ cannot be discarded. For example, the number 35 has two witnesses: 5 and 7. Let us discard the number 5. The number 21 has also two witnesses: 3 and 7. Since the number 7 cannot be discard (otherwise 35 would not have any witness), the number 3 is discarded. However, now the number 15 has not any witness, since 3 and 5 were discarded. The same happens for other numbers and at the final no $w_i$ can be discarded. Thus, calling by $W_{+1}$ ($W_{-1}$) the set of all integer number $n$ having $\mu(n) = +1$ (-1), the set of witnesses chosen is $W = W_{+1}$ if $|W_{+1}| < |W_{-1}|$ otherwise $W = W_{-1}$.

A similar situation is found if one assumes the opposite question: which numbers in the series formed by integer numbers having $\mu \neq 0$ has $\mu = -1$. Once more there will be two possibilities for the witnesses: 1) The set of numbers having $\mu = -1$. In this case the set of witnesses is the same set of numbers to be tested and $U_Q$ is given by (32). 2) The set of numbers having $\mu = +1$. In this case, the operation $U_Q$ is given by (33) and once more none element of the set of witnesses can be discarded. Furthermore, once again $W = W_{+1}$ if $|W_{+1}| < |W_{-1}|$ otherwise $W = W_{-1}$. Hence, in both cases the set of witnesses is the same. This implies that one of them is witness of itself and, hence, it is not compressible. This forces the other set to be incompressible too. In other words, the bit sequence $B$ is incompressible and, hence, maximally random implying that prob[$\mu(n) = +1$] = prob[$\mu(n) = -1$] = ½.

## 5. The zeros of the Remann-zeta function and the prime powers

Initially, let us consider the infinite bit sequence $B = b_1, b_2, b_3,\ldots$ where $b_i = 1$ if $i = p^k$ ($p$ is a prime number and $k$ is an integer) otherwise $b_i = 0$. Therefore, the bit sequence $B$ identifies all prime powers (obviously including the primes when $k = 1$). The bits of $B$ can be produced by the following two deterministic algorithms (their complexities are not relevant here):

$A_1$: Using the first derivative of the Chebyshev function to identify the prime powers:

$$\psi'(x) = 1 - \sum_k x^{\rho_k - 1} - \frac{1}{x(x^2 - 1)}. \tag{38}$$

In (38) $\rho_k$'s are the non-trivial zeros of the Riemman-zeta function. Since the Chebyshev function $\psi(x)$ is a step function that jumps at each prime power $p^k$, its first derivative $\psi'(x)$ is zero everywhere except for spikes at the values of $x$ that are prime powers [13]. Let us rewrite (38) as

$$\psi'(x) = 1 - \sum_k x^{\varepsilon_k - 1} \cos(\ln(x) s_k) - \frac{1}{x(x^2 - 1)}. \tag{39}$$

In (39) as before $\varepsilon_k$ and $s_k$ are, respectively, the real and (positive) imaginary parts of the $k$-th non-trivial zero (the sum of the conjugated zeros was done). Thus, the algorithm to produce $B$ using (39) is as shown:

I) $b_1 = 0$.
II) $b_i = \delta(\psi'(i))$ for $i \neq 1$, where $\delta(y) = 0$ if $y = 0$ otherwise $\delta(y) = 1$.

$A_2$: Using the prime numbers. Having a list of prime numbers, the input integer number $x$ is divided by the primes in the list until a prime factor is found. Since one prime factor $p$ was found out, one can simply divide $x$ by $p$ repeatedly until a zero is obtained as remainder (in this case $x$ is prime power) or a fractional number is obtained as quotient (in this case $x$ is not prime power).

One can note that both algorithms, $A_1$ and $A_2$, even being very different, can be used to provide the same bit sequence $B$. While $A_1$ uses the non-trivial zeros of the Riemann-zeta function as resource, $A_2$ uses the set of prime numbers to produce $B$. It is expected that some properties of $A_1$ and $A_2$ must be included in $B$ and vice-versa. For example, perhaps the randomness of $B$ is related to the randomness of prime numbers. Since $B$, $A_1$ and $A_2$ provide the same information (the numbers that are prime powers), they must be related: it is possible to get $B$ from $A_1$ and $A_2$ (as shown before) and it is also possible to get information of $A_1$ and $A_2$ from $B$. In other words, it is possible to get the primes and the non-trivial zeros of the Riemann-zeta function from $B$. In fact, to obtain the primes from $B$ is a trivial task. For the $n$-th bit 1 in $B$, $n$ is a prime power. An additional test can distinguish primes from prime powers ($k > 1$). Thus, the whole set of primes can be recovered from (the infinite bit sequence) $B$.

To obtain the non-trivial zeros of the Riemann-zeta function from $B$ is a more complicated task. Since all zeros are used to define a single bit of $B$ (Eq. (39)), all bits of $B$ are used to obtain each zero. This is done by inverting (39). In order to have a unique relationship between $B$ and the zeros, the number of unknowns of the zeros and the number of bits of $B$ must be the same. Hence, zeros with different real parts are not allowed. Thus, the imaginary part of the non-trivial zeros ($s_k$) are the solutions of the system:

$$S\left[1 - n^{-1/2}\sum_k \cos(\ln(n)s_k) - \frac{1}{n(n^2-1)}\right] = b_n, \quad n = 2, 3... \qquad (40)$$

The constant value for the real part of the zeros makes the number of unknown equal to the number of equations (equal to the number of bits). A numerical solution of a finite version of (40) can provide an approximation for a finite amount of zeros.

## 6. Conclusions

The RH was discussed using four different points of view:

1) It was shown that making the real part of the non-trivial zeros equal to ½, an indeterminacy of the type 0/0 in the inner product of two coherent states can be avoided.
2) It was shown that it is always possible to construct (at least in principle) a physical system related to any finite amount of zeros on the critical line. Furthermore, the results using the quantum state (12) and the Hermitean operator (17) gives a physical reason for the non-trivial zeros of the Riemann-zeta function to avoid some places out of the critical line. It is important to stress that this result remains valid for any set of zeros, not necessarily consecutive ones.
3) According to the model linking randomness and entanglement used, it was shown that the binary sequence obtained by the application of Möbius function in the sequence of integer numbers that do not have repeated prime factors, is incompressible.
4) It was introduced the binary sequence $B$ in which each bit '1' indicates that its position in the sequence is a prime power. Following, it was presented two algorithms able to produce $B$. The first algorithm, $A_1$, uses the non-trivial zeros of the Riemann-zeta function while the second algorithm, $A_2$, uses the set of prime numbers. Since $B$, $A_1$ and $A_2$ contain the same information, they are the same 'thing' written in three different 'languages'. Thus, the 'translation' between them is possible. The algorithm $A_1$ uses all non-trivial zeros to find each bit of $B$ and all bits of $B$ are required to get the non-trivial zeros. In this case, in order to have a unique solution the real parts of the zeros cannot be variable.

Putting all together, the results here presented reinforce the belief that RH is true.

**Acknowledgements**

This work was supported by the Brazilian agency CNPq via Grant no. 307062/2014-7. Also, this work was performed as part of the Brazilian National Institute of Science and Technology for Quantum Information.